\documentclass[12pt,preprint]{aastex}

\lefthead{Bekki \& Chiba}
\righthead{The origin of the SMC}

\begin{document}

\title{
Formation of  the Small Magellanic Cloud:
ancient major merger as a solution to 
the kinematical differences between old stars and HI gas}
\author{Kenji Bekki} 
\affil{
School of Physics, University of New South Wales, Sydney 2052, Australia}

\and

\author{Masashi Chiba}
\affil{
Astronomical Institute, Tohoku University, Sendai, 980-8578, Japan\\}

\begin{abstract}
 Recent observations of the Small Magellanic Cloud (SMC) have revealed
that
the HI gas shows a significant amount of rotation ($V_{\rm c} \sim 60$
km~s$^{-1}$),
while no or little rotation is evident for the old stellar populations.
We suggest that this unique kinematical difference between these
components
in the SMC can be caused by a major merger event which
occurred in the early stage of the SMC formation. Our simulations show
that
dissipative dwarf-dwarf merging can transform two gas-rich dwarf
irregulars
into
a new dwarf, which consists of a spheroidal stellar component and a
rotating
extended HI disk. The remnant of this dwarf-dwarf merging shows
significantly
different kinematics between stars and gas, in the sense that a gas disk
rotates rapidly while a stellar component shows little rotation.
We thus suggest that the simulated dwarf having
a dynamically hot spheroid and an extended gas disk finally evolves into
the
present SMC after efficient stripping of the outer gas via tidal fields
of
the Galaxy and the Large Magellanic Cloud.
We also suggest that spatial distributions and
kinematics of RGB and AGB stars with different ages in the possible
spheroidal
component of the SMC can provide valuable information on whether and
when
a past major merger event really occurred in the SMC.
\end{abstract}


\section{Introduction}

Recent high-resolution HI observations have revealed
that the SMC has  
a significant amount
of rotation  
with a  circular velocity ($V_{\rm c}$)
of $\sim 60$ km s$^{-1}$  (Stanimirovi\'c et al.  2004, S04).
The observed $V_{\rm c}$ implies that the SMC has a total
mass of $\sim 2.4 \times 10^9 {\rm M}_{\odot}$ within the central 3 kpc
and thus appears to have no dark matter (S04).
The apparent lack of dark matter halo in the SMC
has been suggested to result either from some observational uncertainties 
in estimating the total mass or from very low density of the dark
matter halo (Bekki \& Stanimirovi\'c 2008).

The latest survey of 2046 red giant stars has suggested
that the older stellar components 
of the SMC have a velocity dispersion ($\sigma$)
of $\sim 27.5$ km s$^{-1}$ 
and a  maximum possible rotation
of $\sim 17$ km s$^{-1}$
(Harris \& Zaristky 2006).
This result  is consistent with other kinematical 
studies based on radial velocities of other old and intermediate-age
stellar populations such as PNe and carbon stars
(e.g.,  Dopita et al. 1985; Suntzeff et al. 1986; 
Hatzidimitriou et al. 1997), which implies that
the older stellar component is  a spheroid
that is primarily supported by its velocity dispersion.
These observations and HI ones (e.g., S04) thus suggest
that there is a remarkable difference in kinematics
between older stellar populations and HI gas in the SMC.

Bekki \& Chiba (2008) pointed out that the rotating
gas disk of the SMC can be gradually formed via gas accretion
{\it after the formation of the older spheroidal component.} 
They also pointed out that directions
of intrinsic spin axes of older stars  
and gas in the SMC 
could be  significantly  different to each other, like
polar-ring galaxies.
They did not however discuss how the dynamically hot stellar spheroid
composed of older stars in the SMC can be formed.
If the SMC really has a dynamically hot spheroid composed mostly
of older stellar populations,
a key question is why the SMC has {\it both} a dynamically hot
stellar spheroid and  
a rotating gas disk.

The purpose of this {\it Letter} is to suggest 
a new possible scenario that the SMC could have experienced a major
merger event long time ago in which both the older stellar
spheroid and the rotating gas disk were created.
Based on numerical simulations,
we show that ancient dwarf-dwarf merging can transform
two gas-rich dwarf disks  into one gas-rich spheroid
with an extended gas disk with rotation.
We explain why the SMC {\it as a  (dwarf) spheroidal galaxy}  
could have  a larger amount of HI gas ($>10^9 {\rm M}_{\odot}$)
in the context of the ancient major merger event.
We also discuss advantages and disadvantages of this scenario 
in explaining the observed structural and kinematical
properties of stars with different ages 
in the present SMC.

Although recent extensive and systematic numerical simulations
have investigated physical properties of remnants of 
mergers between luminous disk galaxies
(e.g., Naab et al. 2006; Di Matteo et al. 2007),
they did not investigate models in which {\it merger progenitors
are low-mass dwarfs and  have very extended HI  disks}.
The extended HI  disks are one of the characteristics of 
gas-rich dwarfs (e.g., Hunter 1997; Warren et al 2004) and  modeled
properly by the present study for the first time.
The extended HI disks of merger progenitors 
can play a key role in forming extended gas
disks surrounding old spheroids
in merger remnants, as shown later in this paper.

\section{The model for the merger scenario}

We investigate chemodynamical evolution of gas-rich mergers
between dIs with stellar and gaseous disks embedded in massive
dark matter halos by using our original
chemodynamical  codes (Bekki \& Chiba 2005, BC05)
that can be run on GRAPE systems (Sugimoto et al.1990).
Since the details of numerical techniques and methods (e.g., ways to model
chemical enrichment processes) are already given in  
the above papers, we briefly describe them in the present paper.
We adopt the Burkert profile (Burkert 1995) for the radial
density profile of the dark matter halo of a dI,
because it  can be consistent with rotation curve profiles
of dIs (Burkert 1995).
The total masses of the dark matter ($M_{\rm dm}$)
and the baryonic component ($M_{\rm b}$)
for the dI
are set to be $8.0 \times 10^9 {\rm M}_{\odot}$
and  $2.0 \times 10^9 {\rm M}_{\odot}$, respectively.
The dark matter halo has a large core radius ($r_{\rm c}$)
of 3.5 kpc and the truncation radius of $ 3.4 r_{\rm c}$
(Burkert 1995).

The dI is described as a pure disk  (without a  bulge)
with the initial size of 5 kpc
and the radial ($R$) and vertical ($Z$) density profiles
starting from a  thin  disk
are  assumed to be
proportional to $\exp (-R/R_{0}) $ with scale length $R_{0}$ = 1 kpc 
and to  ${\rm sech}^2 (Z/Z_{0})$ with scale length $Z_{0}$ = $0.2R_{0}$,
respectively.
The HI diameters of gas-rich galaxies are generally observed to be
larger
than their optical disks (Broeils \& van Woerden 1994). A small fraction
of low luminosity galaxies have HI gas envelopes extending out to 4--7
optical radii  (e.g., Hunter 1997) with HI mass to light ratios up to $\sim
20 {\rm M_{\odot}\,L_{\odot}^{-1}}$ (Warren et al. 2004).
Observational studies showed that there are threshold gas densities
for galaxy-scale star formation in disk galaxies (Kennicutt 1998).
Guided by these   observations, 
we assume that the disk can contain only gas for $R>R_{\rm th}$,
where $R_{\rm th}$ is a parameter for the stellar disk size,
owing to the lower gas density for  $R>R_{\rm th}$.
The fraction of gas mass ($M_{\rm g}$) to the sum of
stellar ($M_{\rm s}$) and gaseous ones
is represented by $f_{\rm g}$ and assumed to be a free parameter.
The size of gas disk ($R_{\rm g}$) is set to be four times larger
than that of stellar one ($R_{\rm s}$) in the present study,
which is consistent with those of some dIrr's 
(e.g., Hunter 1997).

Star formation
is modeled by converting  the collisional
gas particles
into  collisionless new stellar particles according to the algorithm
of star formation  described below.
We adopt the Schmidt law (Schmidt 1959)
with exponent $\gamma$ = 1.5 (1.0  $ < $  $\gamma$
$ < $ 2.0, Kennicutt 1998) as the controlling
parameter of the rate of star formation.
The stars formed from gas are called ``new stars'' 
whereas stars initially within a disk  are called ``old stars''
throughout this paper.
Chemical enrichment through star formation and supernova feedback
are assumed to proceed both locally and instantaneously in the present study.
The values of chemical yield and return parameter
are 0.004 and 0.3, respectively,
and the initial gaseous metallicity in the dI is ${\rm [Fe/H]} = -1.0$.
The total masses of old and new stars  are referred
to as $M_{\rm os}$ and $M_{\rm ns}$, respectively.

The mass ratio of the two merging dIs ($m_2$), 
the pericenter distance ($r_{\rm p}$),
and the eccentricity ($e_{\rm p}$) are 
assumed to be free parameters.
The orbit of the two dIs is
set to be inclined by 45 degrees
with respect to  the $xy$ plane and the distance between
the center of mass of the two dIs
is  20 kpc.
The spin of each galaxy in a merger
is specified by two angles $\theta_{i}$ and
$\phi_{i}$, where suffix  $i$ is used to identify each galaxy.
$\theta_{i}$ is the angle between the $z$ axis and the vector of
the angular momentum of a disk.
$\phi_{i}$ is the azimuthal angle measured from the $x$ axis to
the projection of the angular momentum vector of a disk onto the $xy$
plane.

Although we run many models with different parameters,
we show only one model with  
$f_{\rm g} =  0.7$, $R_{\rm th}=1.25$ kpc
(i.e., $R_{\rm g} = 4R_{\rm s}$),
$m_{2}=0.5$, $r_{\rm p}=2$ kpc, $e_{\rm p}=1.0$, 
$\theta_{1}=30^{\circ}$,
 $\theta_{2}=120^{\circ}$,  $\phi_{1}=90^{\circ}$,
$\phi_{2}=30^{\circ}$.
This is mainly because the merger remnant of the model shows more
clearly both kinematical differences between gas and stars and
extended gas disk with rotation.
The simulations have mass and size resolutions of $10^4 {\rm M}_{\odot}$
and 90 pc, respectively, and both the GRAPE-SPH code  
(Bekki \& Chiba 2006) and
the one adopted  in BC05  are  used for each model. 
We show the results of the model based on the same code as BC05 
and will give the results based on GRAPE-SPH codes 
in our forthcoming papers (Bekki \& Chiba 2008) in order
to describe in detail 
the dependences of the results on modeling of the interstellar medium.

One of the new ingredients in the present simulation is that the merger
progenitor dIs have very extended gas disks that are consistent
with observations (e.g., Hunter 1997; Warren et al. 2004).
We confirm that the  extended  gas disks in merger progenitor dIs
play a vital role in forming the extended gas disk surrounding
old stars in the merger remnants irrespective of 
initial gas mass fractions. This importance of extended gas disks
was not investigated in previous simulations (e.g., Naab et al. 2006).
In the present study,
we consider that the dI-dI  merging leading to
the formation of the SMC occurred long before the strong LMC-SMC
interaction commenced about $3-4$ Gyr ago (BC05): possibly
this merger event might have occurred far away from the Galaxy
in order to have the low relative velocity between two merging dIs.
We consider that the low relative velocity would be possible,
because the two dIs were either initially a pair
or in a very small group (of galaxies) with a smaller circular velocity
that merged the outer region of the Galaxy's halo long time ago.

\section{Results}

Figure 1 describes how dwarf-dwarf merging can transform two dIs
into a new dwarf with a central spheroid and an extended gas  disk.
Owing to strong violent relaxation in the central region
of the merger, the inner stellar disks are completely destroyed and  form
a slightly flattened spheroidal component with a half-mass radius of 2.0 kpc.
Although the gas disk of the larger dI
can be temporarily  disturbed strongly  by the merging,
it finally becomes  a new extended gas disk
after dissipative merging with that of the smaller dI.
The star formation rate can reach the maximum value
of  $2.4 {\rm M}_{\odot}$ yr$^{-1}$
at $T=0.6$ Gyr during the merging
owing to efficient and rapid  transfer of gas toward
the central region of the merger.
New stars formed during merging also show a significantly
flattened spheroidal distribution
with a half-mass radius of 1.3 kpc.
The outer low-density part of the stellar spheroid 
composed of metal-poor stars ($-0.9 < {\rm [Fe/H]} < -1.0$)
might well be identified as a stellar halo 
around  the remnant.

Figure 2 shows that the final total mass of the remnant 
within the central 3 kpc at $T=1.7$ Gyr
is about $2.8  \times  10^9 {\rm M}_{\odot}$,
which is consistent with the observation by S04.
However, the total stellar  mass 
(i.e., $M_{\rm s}=M_{\rm os}+M_{\rm ns}$) within the central
3 kpc is $0.4 \times 10^9 {\rm M}_{\odot}$
and thus appears to be significantly  smaller than the observed one
($M_{\rm s}=1.8 \times 10^9 {\rm M}_{\odot}$)
of the present SMC 
with the assumed mass-to-light-ratio of $\sim 1$ (S04). 
The total mass of the remaining gas is
$1.6 \times 10^9 {\rm M}_{\odot}$ 
and most of the gas is located at $R>1$ kpc. 
The gas can be used for further star formation in the remnant
so that the total stellar mass within the central 
3kpc can dramatically increase
within several Gyrs.

Figure 3 shows how radial profiles of 
line-of-sight velocities ($V$) and velocity dispersions ($\sigma$)
are different between stellar and gaseous components.
The gaseous component clearly shows a large amount
of rotation with the maximum
$V$ of $59$ km s$^{-1}$ and a small central velocity dispersion
of $\sigma = 24$ km s$^{-1}$ (i.e., $V/\sigma \sim 2.5$) 
if the remnant is seen almost from  the edge-on.
The stellar component, on the other hand,
shows a smaller amount of  rotation of $V\sim 20$ km s$^{-1}$ and a larger
maximum velocity dispersion of 
$\sigma \sim 48$ km s$^{-1}$ (i.e., $V/ \sigma \sim 0.4$).
The simulated high and low  $V/\sigma$ in gas and stars, respectively,
are consistent with the observations of the SMC 
(S04; Harris \& Zaristky 2006), 
which means that the model  can reproduce well
the present SMC with a possible spheroid with dynamically hot kinematics 
and an extended gas disk with rotation.

Figure 4 shows that there is a negative  metallicity gradient
within the central 2 kpc
($\Delta {\rm [Fe/H]} / \Delta R \sim -0.05$  dex kpc$^{-1}$)
for the gaseous component in the sense that
the inner part is more metal-rich. The outer part 
of the remnant ($R>3$ kpc), which is composed mostly 
of gas ($f_{\rm g}>0.6$), shows ${\rm [Fe/H]} < -0.95$
owing to severe suppression of star formation and the resultant
much less efficient chemical enrichment.
These results imply that if the outer gas disk of the remnant
is tidally stripped by other giant galaxies, the gaseous 
stream can be very metal-poor.
The new stars also show a negative yet weak  metallicity gradient
($\Delta {\rm [Fe/H]} / \Delta R \sim -0.03$  dex kpc$^{-1}$)
as a result of inward transfer of gas chemically polluted 
by supernova. Thus the present 
chemodynamical simulations demonstrate that
dissipative dwarf-dwarf merging can transform two dIs into
one dwarf that has both  a stellar spheroid with younger, 
more metal-rich stellar population in its inner part
and an extended, more  metal-poor outer gas disk.

\section{Discussion and conclusions}

Although previous numerical simulations of 
galaxy merging with large gas mass fractions
and unique orbital configurations
have already reproduced  dynamically hot stellar spheroids
with extended  gas disks or polar-rings (Bekki 1998),
the present study has first shown the formation of stellar spheroids
with extended gas disks with rotation from 
gas-rich {\it dwarf-dwarf merging.}
We confirm that the formation processes of dwarf spheroidal  with extended
gas disks do not depend so strongly on model parameters as long as
the merger precursor dI has  an extended gas disk:
the size ratios of stellar disks to gaseous ones in dIs 
can be a key parameter that controls physical properties of the remnants
of dwarf-dwarf merging.

The present study has first suggested   a scenario in which  
both the stellar spheroid and the extended HI gas disk
with rotation in the SMC  could have been formed in
an ancient dissipative dwarf-dwarf merger event. 
The observed hot kinematics of older 
stellar populations can be thus due to violent dynamical relaxation
associated with the merging. 
The observed kinematical differences between older stellar and gaseous
components in the SMC 
result from differences in dynamical evolution between the two
components
during the merging.
In this scenario,
the SMC with an extended HI disk 
interacted strongly with
the LMC and the Galaxy from  $2-3$ to $0-0.2$ Gyrs ago
(e.g.,  Bekki \& Chiba 2007)
so that it could  lose the gas disk and
then finally  can become the present SMC: the merging
happened much earlier than the recent LMC-SMC-Galaxy interaction.
The stripped HI gas is now observed to be either the Magellanic
Stream (MS) or the Magellanic Bridge (MB).

If this scenario is correct,
then when did the SMC experience such a dwarf-dwarf merger
event ?
Since stellar populations formed before the merger event
should have dynamically hot kinematics in this scenario,
the youngest age of stellar populations that show {\it both} spheroidal
distributions and no or little rotation can correspond
to the epoch when the merging occurred.  
Recent observations of AGB stars in the SMC have 
reported that (i) the average age of the old and intermediate
stellar populations is $7-9$ Gyr and (ii)
the  stars have a more  regular distribution
and appear  to be a slightly flattened ellipsoid
(e.g., Cioni et al. 2000; 2006).
However, owing to the lack of observations
on  dependences of kinematical properties for  AGB/RGB stars
on their ages,
it is currently difficult to derive the youngest age
of stars that consist of the possibly dynamically hot spheroid.

Previous models for the formation of 
the MS and the MB
suggested that the distribution of the  gaseous component in the SMC
is required to be significantly more extended than
that of the stellar one 
(e.g., Yoshizawa \& Noguchi 2003; Muller \& Bekki 2007).
The present scenario naturally explains why the required
distribution is possible in the SMC.
However, the scenario also predicts that
the merger remnant (i.e., the SMC)
can have stars with the total mass of about $10^8 {\rm M}_{\odot}$ 
for $3 < R < 5$ kpc.
This means that the MS and MB can possibly contain
stars, if they are formed from tidal stripping,
which is more effective in the outer regions of the SMC.
The SMC with actively star-forming central regions 
might well look like 
a blue compact dwarf (BCD),
if it is observed at the epoch when the dwarf-dwarf  merging
is in the very late stage or just completed (i.e., a high redshift
universe).
We thus  suggest that gas-rich dwarf-dwarf merging
can  provide a possible  evolutionary link between
BCDs and dIs with older stellar spheroids (Bekki \& Chiba 2008).

Recent observations have reported that some dwarf dIs 
have old or intermediate-age stellar populations
in their halo regions (e.g., Battinelli \& Demers 2006;  
Battinelli et al. 2007;
Battinelli  et al. 2007;
Demers et al. 2006).  
Some ``transition'' dwarfs,
which are intermediate-class of objects
between  dIs and dEs/dSphs,
are observed to have early-type outer structures
and young stellar populations in their centers 
(e.g., Dellenbusch et al. 2008).
Some dwarfs with older stellar spheroids (e.g., NGC 404 and NGC 6822)
are observed to have  extended HI gas  disks
(e.g., del R\'io et al. 2004; Demers et al. 2006).
Therefore the SMC is not 
a rare example of dIs that have older stellar spheroids:
understanding the origin of the SMC could help us to better
understand the formation and evolution of dwarfs listed above.

Irregular optical appearances of dIs are simply due to inhomogeneous
distributions of star-forming regions in {\it gas disks}
and therefore do not  provide information on spatial
distributions of underlying older stellar populations.
The present study suggests that
some fraction of dIs with outer spheroids
can be merger remnants.
Although recent observations have revealed that
many dIs have extended low surface brightness structures
(e.g., Minniti \& Zijlstra 1996; Aparicio \& Tikhonov 2000),
it is observationally  unclear 
how much fraction of dIs have {\it spheroids}  
composed mostly
of older stellar populations: the number fraction
dIs possibly formed from merging is unclear.
It could  be possible that {\it some}  of the present dSphs were previously
dIs with older spheroids formed by dwarf-dwarf
merging and later became dSphs after losing 
HI gas  via tidal and ram pressure stripping.
Future  numerical simulations of dwarf-dwarf merging
with a wide rage of model parameters need to  discuss
this possibility of dSph formation at very  high-z.

\acknowledgments
We are  grateful to the anonymous referee for valuable comments,
which contribute to improve the present paper.
K.B. acknowledges the financial support of the Australian Research
Council throughout the course of this work.
The numerical simulations reported here were carried out on GRAPE
systems kindly made available by the Center for Computational
Astrophysics (CfCA)
at National Astronomical Observatory of Japan (NAOJ).
This work was financially supported by CfCA.

%%%%%%%%%%%%%%%%%%%%%%% Figure Captions
\newpage

\begin{figure}
\plotone{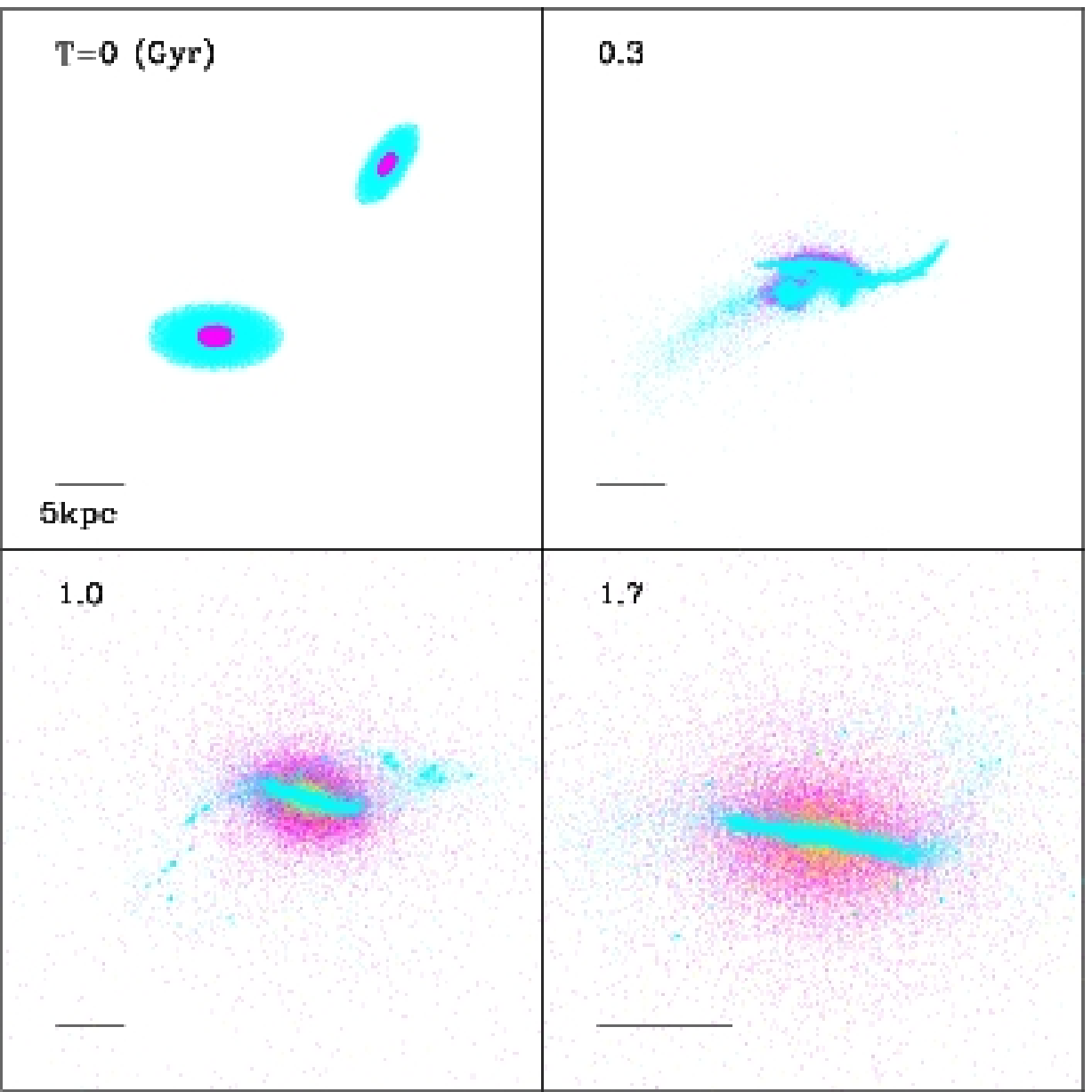}
\figcaption{
Mass distributions of old stars (magenta),
gas (cyan), and new stars (yellow)
of the unequal-mass dI-dI  merger with $m_2=0.5$
projected onto the $x$-$z$ plane at four different time steps.
The time $T$ in units of Gyr is shown 
in the upper left corner of each panel.
Frames measure 40 kpc in the first three
and 20 kpc in the last so that the extended disk can be more clearly
seen at $T=1.7$ Gyr. 
Only  old stars can be seen  for the central 1.25 kpc of two dIs
in the first frame, because the stellar particles are overlaid  on
gaseous ones:
gas particles exist
in the central 1.25 kpc.
\label{fig-1}}
\end{figure}

\newpage
\begin{figure}
\plotone{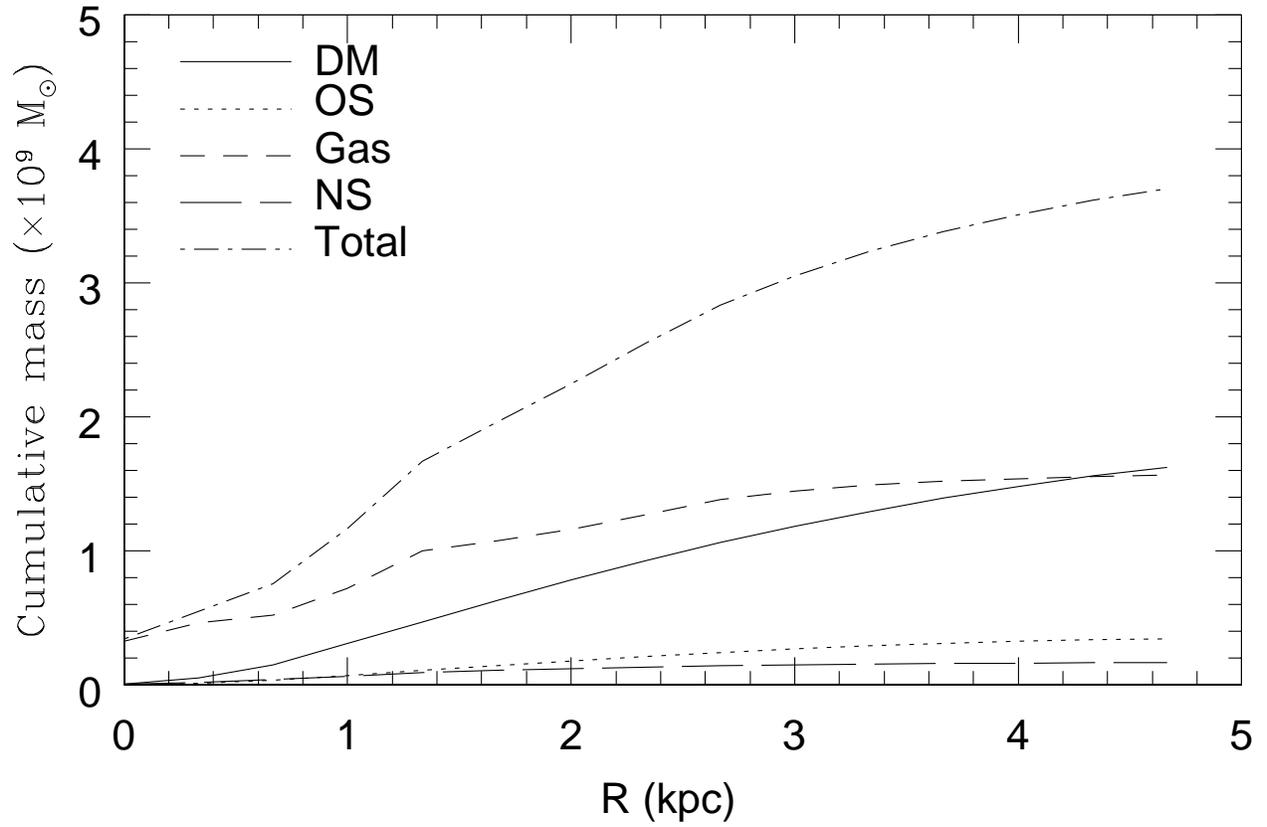}
\figcaption{
The total  mass within $R$ for dark matter halo 
(solid, referred to as DM),
old stars (dotted, OS), gas (short-dashed), new stars (long-dashed),
and total (dash-dotted) in the merger remnant.
\label{fig-2}}
\end{figure}

\newpage
\begin{figure}
\plotone{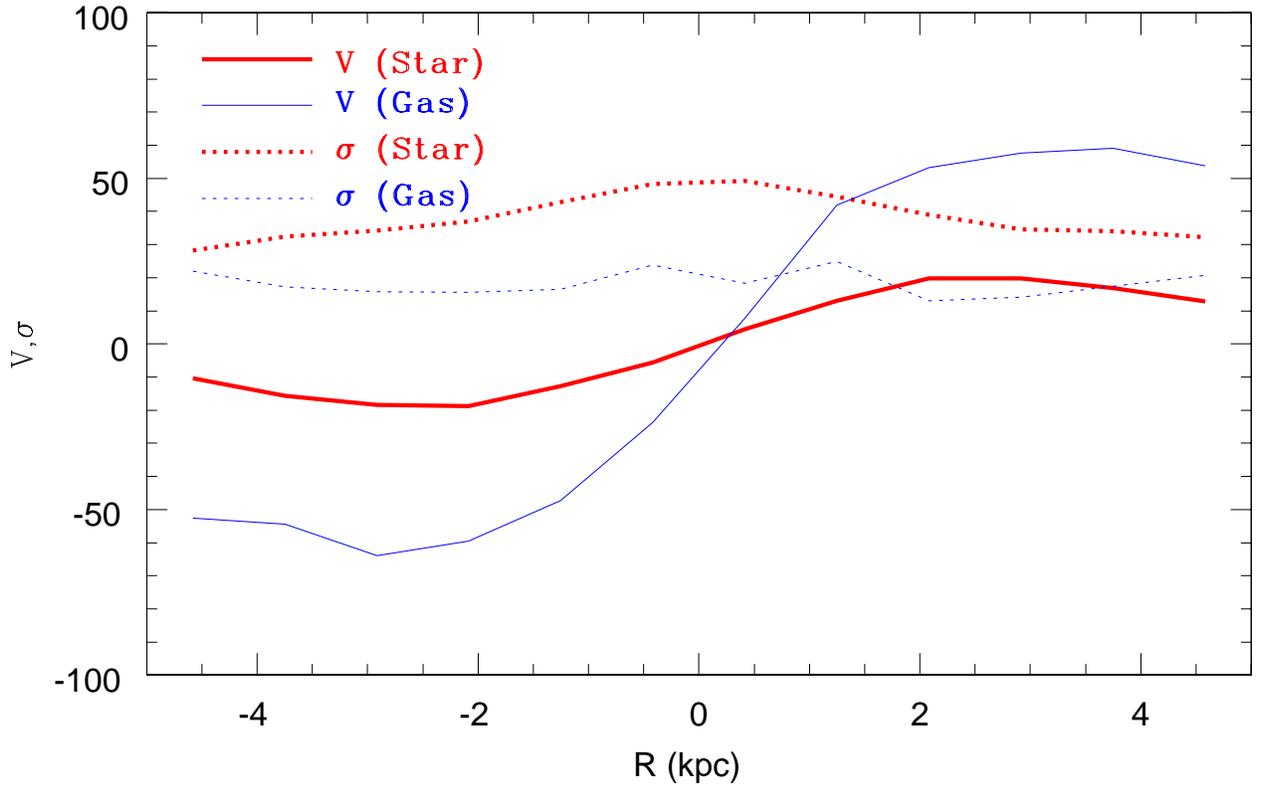}
\figcaption{
The radial dependences of line-of-sight velocities ($V$) 
for stars (red, thick solid) and gas (blue, thin solid) and dispersions
($\sigma$)  for stars  (red, thick dotted) and gas (blue, thin dotted)
in the merger remnant.
Here the remnant is seen from the $x$-axis so that the $y$-components
of velocities of particles are used for the radial $V$ and $\sigma$ profiles.
\label{fig-3}}
\end{figure}

\newpage
\begin{figure}
\plotone{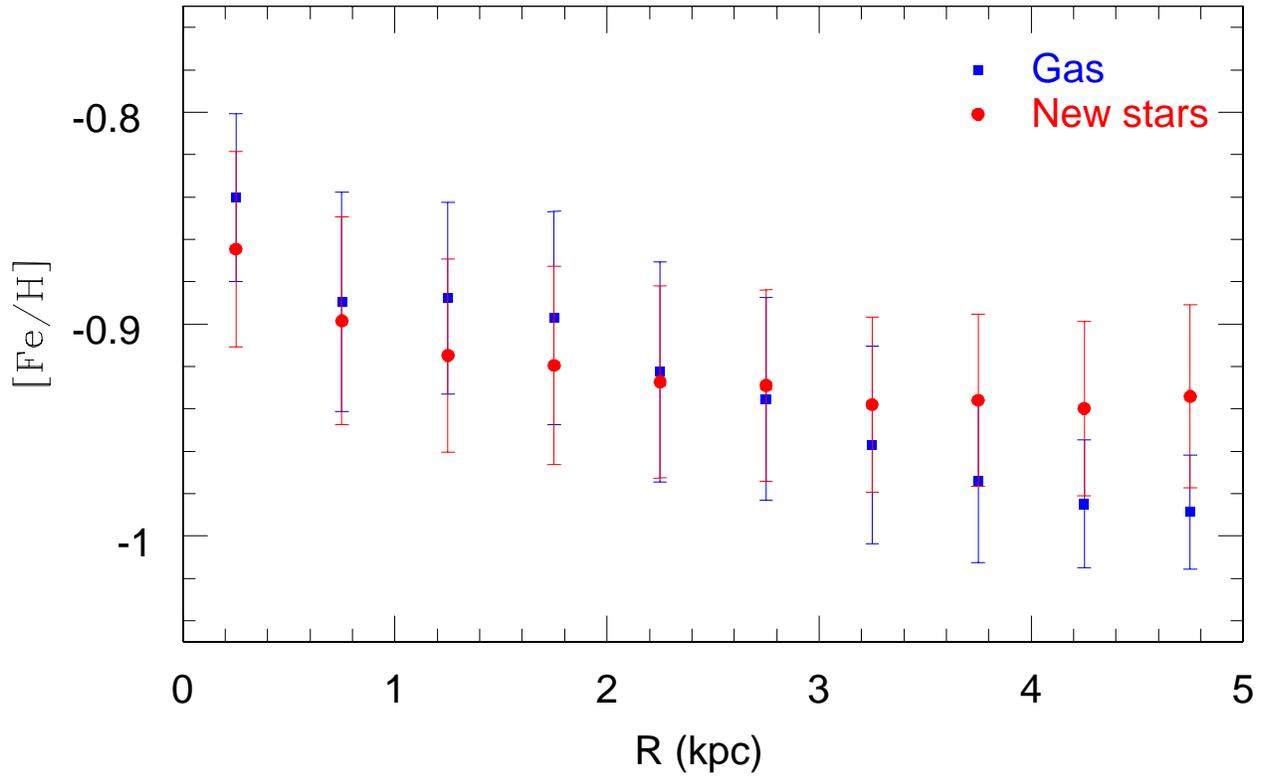}
\figcaption{
Metallicity gradients of gas (squares, blue) and new stars (circles, red)
in the merger remnant for the model with
the initial gaseous metallicity of ${\rm [Fe/H] } =  -1$.  Note that the outer
regions ($R>3$ kpc) of the  gas disk show metallicities
similar to the original ones (i.e., ${\rm [Fe/H]} =-1$).
\label{fig-4}}
\end{figure}

\end{document}